\documentclass[12pt]{article}
\usepackage{amssymb}
\usepackage{graphicx}
\begin{document}
\begin{titlepage}
\hskip 11cm \vbox{
\hbox{Bicocca--FT--00--01}
\hbox{IFUP--TH 2000-10}}
\vskip 1cm
\centerline{\bf Topological susceptibility in full QCD at zero and
finite temperature}
\vskip 0.8cm
\centerline{B.~All\'es$^{\rm a}$, M.~D'Elia$^{\rm b}$ and A.~Di~Giacomo$^{\rm b}$}
\vskip .3cm
\centerline{\sl $^{\rm a}$ Dipartimento di Fisica, Universit\`a di 
  Milano--Bicocca and}
\centerline{\sl INFN--Sezione di Milano, I--20133 Milano, Italy}
\vskip .1cm
\centerline{\sl $^{\rm b}$ Dipartimento di Fisica, Universit\`a
di Pisa and INFN--Sezione di Pisa, I--56127 Pisa, Italy}
\vskip .1cm
\vskip 1cm
\begin{abstract}
We present a study of the topological susceptibility $\chi$ 
on the lattice for full QCD with 2 and 4 flavours of
staggered fermions at zero and finite temperature $T$.
We find that $\chi$ presents a sharp drop across the deconfinement
transition. We also study the dependence of $\chi$ on the quark mass
at $T=0$: we have no conclusive evidence for the expected chiral
behaviour. 
\end{abstract}
\vfill

\vskip.3cm
\end{titlepage}
\eject

\section{Introduction}
\label{sec:intro}

The topological susceptibility $\chi$ plays a relevant role in
understanding several low energy properties of QCD.
It is defined as the zero--momentum two--point correlation function of
the topological charge density~$Q(x)$~\cite{witten,veneziano},
\begin{eqnarray}
 \chi &\equiv& \int \hbox{d}^4x \;\; \partial_\mu \; 
              \langle 0| \hbox{T}\{K^\mu(x) Q(0)\}|0\rangle\, , \nonumber \\
 Q(x) &\equiv& {g^2\over 64\,\pi^2} \; \epsilon^{\mu\nu\rho\sigma}\;
              F^a_{\mu\nu}(x) F^a_{\rho\sigma}(x) \; ,
\label{eq:chiq}
\end{eqnarray}
where $K^\mu(x)$ is the Chern current 
\begin{equation}
 K^\mu(x) \equiv {g^2 \over 16\,\pi^2} \;\epsilon^{\mu\nu\rho\sigma}\;
                 A^a_\nu(x) \; \left(\partial_\rho A^a_\sigma(x) - {1\over
                 3} \; g \; f^{abc} A^b_\rho(x) A^c_\sigma(x)\right),
\label{eq:kmu}
\end{equation}
which satisfies $Q(x)=\partial_\mu\,K^\mu(x)$. In these expressions
$g$ is the QCD coupling constant and $f^{abc}$ are the structure
functions of SU(3). In Eq.~(\ref{eq:chiq}) we keep the 
derivative out of the expectation value in the definition of 
$\chi$ according to the continuum prescription~\cite{crewther}.

The topological susceptibility in the quenched theory at zero
temperature has been widely studied. 
In particular, a non--zero value of $\chi$ provides an
explanation for the large mass of the $\eta'$ particle, the would--be
Goldstone boson of the axial symmetry~\cite{witten}. 
In Ref.~\cite{npb494} we have found
that the value of $\chi$ in pure Yang--Mills theory is 
$\chi=(175(5) \hbox{ MeV})^4$ in agreement with the
phenomenological expectation~\cite{veneziano}. For a review of other
calculations see~\cite{teper} and references therein.

In full QCD, Ward identities and current algebra relations
imply in the chiral limit~\cite{veneziano,leutwyler}
\begin{equation}
 \chi = {m \over N_f} \; \langle \overline{\psi} \psi \rangle \; ,
\label{eq:psibarpsi}
\end{equation}
where $m$ is the fermion mass and $N_f$ the number of fermions. It is
interesting to verify on the lattice the linear dependence on the mass
displayed by Eq.~(\ref{eq:psibarpsi}). 

The topological susceptibility at non--zero temperature for pure
Yang--Mills theory has also been studied. In Ref.~\cite{npb494} it has been
shown that it is rather constant versus $T$ below the deconfinement
temperature $T_c$ and it undergoes a sharp drop across the phase
transition. It is also interesting to study the
behaviour of $\chi$ for full QCD as a function of $T$.
In particular, Eq.~(\ref{eq:psibarpsi}) suggests the existence of a
drop in the signal of $\chi$ across $T_c$ also for the full QCD
case~\cite{leutwyler}, (see also~\cite{shuryak}).

In the present paper we will present a lattice determination of 
$\chi$ in full QCD for $N_f=2$ degenerate flavours of staggered 
fermions at finite and zero temperature and for $N_f=4$ at finite temperature. 
The Monte Carlo simulation was done on the APE--Quadrics in Milano and Pisa. 

The method used to extract $\chi$ from the lattice
is described in Section~\ref{sec:operators}.
The results and technical details of the simulation for the $N_f=4$
and $N_f=2$ cases are shown in Section~\ref{sec:4} and 
Section~\ref{sec:2} respectively. The conclusions are drawn in the
last section.

\section{Method to determine $\chi$}
\label{sec:operators}

We have simulated the theory on a space--time lattice.
The topological charge density has been measured by making use of the 
lattice operator
\begin{equation}
Q_L(x)\equiv -{1\over 2^9\pi^2}\; \sum_{\mu\nu\rho\sigma = \pm 1}^{\pm 4}
 {\widetilde\epsilon}_{\mu\nu\rho\sigma}
 \hbox{Tr}\, \left(\Pi_{\mu\nu}(x)\;\Pi_{\rho\sigma}(x)\right)\;,
\label{eq:ql}
\end{equation}
where $\Pi_{\mu\nu}(x)$ is a plaquette in the $\mu$--$\nu$ plane and
${\widetilde\epsilon}_{\mu\nu\rho\sigma}$ is a generalized
Levi--Civita tensor which acquires an extra minus sign
for each of its indices going negative. Green functions containing
$Q_L(x)$ differ from the continuum counterparts by a finite
renormalization $Z$~\cite{campostrini}, in short
\begin{equation}
Q_L(x) = Z \, a^4\, Q(x) + O(a^6)\;,
\label{eq:z}
\end{equation}
where $a$ is the lattice spacing. Actually $Q_L(x)$ mixes
with fermionic operators during renormalization but it has been shown
that the off--diagonal mixings are negligible~\cite{plb350}.

The lattice topological susceptibility $\chi_L$ is defined as 
\begin{equation}
\chi_L\equiv{\langle Q^2_L\rangle\over V}\; ,
\label{eq:chil}
\end{equation}
where $V$ is the lattice space--time volume and $Q_L$ is the total
topological charge, $Q_L\equiv\sum_x Q_L(x)$. The definition of
$\chi_L$ differs from the continuum expression for $\chi$ in
Eq.~(\ref{eq:chiq}) by contact terms which must be
subtracted. The relationship between $\chi$ and $\chi_L$
is
\begin{equation}
 \chi_L = a^4\, Z^2\, \chi + M\;,
\label{eq:chilzm}
\end{equation}
where $M$ contains mixings with operators with compatible quantum
numbers. 

We have measured $Q_L(x)$ 
after having applied two smearing steps with smearing parameter 
$c=0.9$~\cite{christou}. The smearing process 
allows the operator to be less sensitive to the quantum fluctuations.
Therefore the renormalization constant $Z$ approaches 1 and $M$
approaches 0. As a consequence
the statistical errors are strongly diminished~\cite{christou}. 

We evaluate $Z$ and $M$ by using the ``heating method'' which provides 
a non--perturba-\break tive determination 
of the renormalization constants~\cite{vicari}.
To calculate $Z$, several steps of an updating algorithm (the same 
used during the Monte Carlo simulation) are applied on a configuration
containing 
a charge +1 classical instanton. These updatings thermalize first
the quantum fluctuations, responsible for the renormalization effects,
and due to the slowing down leave the topological content unchanged.
The measurement of $Q_L$ on such updated configurations yields $Z \cdot Q$. As
$Q$ is known, one can extract $Z$. Notice that this procedure is equivalent
to imposing the continuum value for the 1--instanton charge (in the
$\overline{\rm MS}$ scheme it is +1) and extracting the finite multiplicative
renormalization $Z$ by evaluating a matrix element of $Q_L$.

The additive renormalization $M$ is obtained in a similar way. We
apply a few heating steps with the same updating algorithm as above 
on a zero--field configuration. The topological susceptibility is then
measured. This provides the value of $M$, if no instantons have been
created during the few updating steps. 
As explained in~\cite{npb494}, cooling tests must be done to check
that the background topological charge has not been changed during the
heating. Configurations where the topological content has been changed
are discarded from the sample.

Once $Z$ and $M$ are known, we can extract $a^4\chi$ from
Eq.~(\ref{eq:chilzm}). We have fixed $a$ in physical units by
measuring the string tension $\sigma$ on a $16^4$ lattice and assuming that
$\sqrt{\sigma}=420$ MeV. To this purpose,
Wilson loops ranging from $1\times 1$ to $8\times 8$ 
have been evaluated by using smeared spatial links and a
best fit has been performed for the interquark potential to the form
$V(r)=V_0 + a_0/r + \sigma\, r$~\cite{bali}.

\section{Full QCD with $N_f=4$}
\label{sec:4}

We have numerically simulated four flavours of quarks by using
staggered fermions of bare mass $am=0.05$. For the pure gauge sector
of the theory the Wilson action has been chosen. The $\Phi$--type HMC
algorithm~\cite{gottlieb} has been used for the updating.
We have simulated at the following values of
$\beta\equiv6/g^2$=5.00, 5.02, 5.04, 5.05 5.06 and 5.10 ($g$ is the lattice
bare coupling).

\begin{figure}[tbh]
\begin{center}
\includegraphics[width=0.5\textwidth]{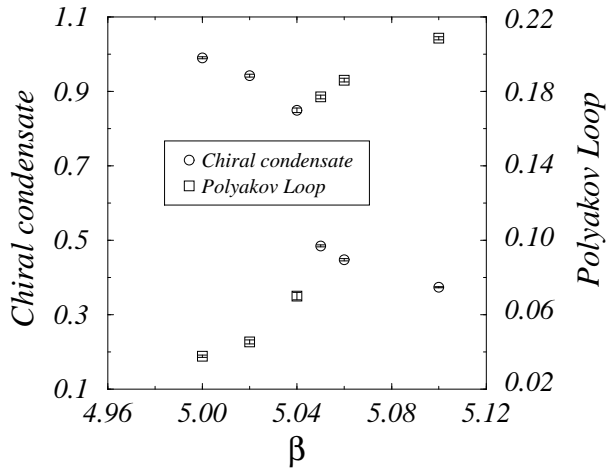}
\end{center}
\caption{\small{Polyakov loop and chiral condensate at finite $T$ for $N_f=4$.}}
\label{fig:4poly}
\end{figure}

The lattice volume was $16^3\times 4$. According to Ref.~\cite{brown}, 
at this temporal size 
the deconfinement transition occurs at $\beta_c=5.04$. We have checked
this value by computing the Polyakov loop and the chiral condensate as
shown in Fig.~\ref{fig:4poly}.

In the asymmetric lattice the temperature has been determined 
by the definition $T=~1/(a\,L_t)$, where $L_t$ is the temporal size.

\begin{figure}[tbh]
\begin{center}
\includegraphics[width=0.5\textwidth]{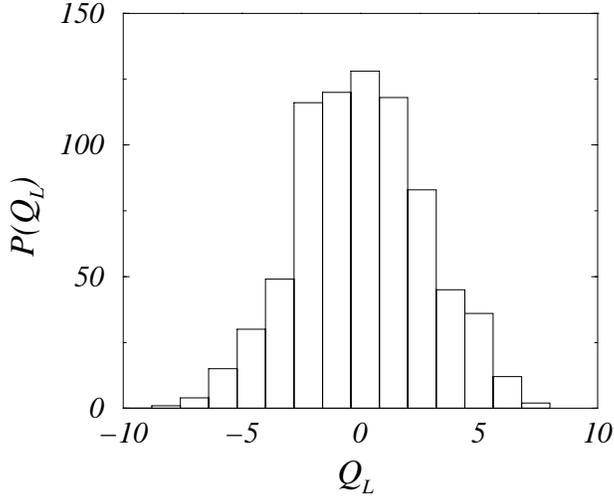}
\end{center}
\caption{\small{Distribution of topological charge $P(Q_L)$ at $\beta=5.04$.}}
\label{fig:4histo}
\end{figure}

We have checked that the distribution
of topological charge is thermalized enough. In Fig.~\ref{fig:4histo} we
show such a distribution for $\beta=5.04$ after 30 cooling steps. 
At smaller values of $am$ ($am \simeq 0.01$) and higher $\beta$'s, simulations
are affected by a slowing down in the sampling~\cite{plb389,schilling}.

In Table~{\ref{tab:nf4}} we list the numerical results.
The lattice spacing $a$ has been determined at
$\beta=$5.00, 5.04 and 5.10 on a $16^4$ lattice. For the other
values of $\beta$, it was extrapolated by splines. From the 
value at $\beta_c$, $a=0.30(2)$ fm, we infer the critical
temperature to be $T_c=164(11)$ MeV.

\begin{table}[hbtp]
\setlength{\tabcolsep}{1.88pc}
\centering
\caption{$\chi$, $T/T_c$ and $a$ versus $\beta$
for $N_f=4$ on a $16^3\times 4$ lattice.}
\begin{tabular}{cccll}
\hline
\hline
$\beta$ & $a$/fm & $T/T_c$ & $10^{-8}\times\chi$/MeV$^4$ \\
\hline
5.00 & 0.31(4) & 0.9677 & 1.61(43) \\
5.02 & 0.306$\;^\dagger$ & 0.9804 & 1.13(30) \\
5.04 & 0.30(2) & 1.0000 & 1.21(30) \\
5.05 & 0.296$\;^\dagger$ & 1.0126 & 0.29(15) \\
5.06 & 0.292$\;^\dagger$ & 1.0274 & 0.16(12) \\
5.10 & 0.270(8) & 1.1110 & 0.04(12) \\
\hline
\hline
\noindent{$\rm _{\dagger\; Extrapolated,\; see\; text.}$}
\label{tab:nf4}
\end{tabular}
\end{table}

\begin{figure}[tbh]
\begin{center}
\includegraphics[width=0.5\textwidth]{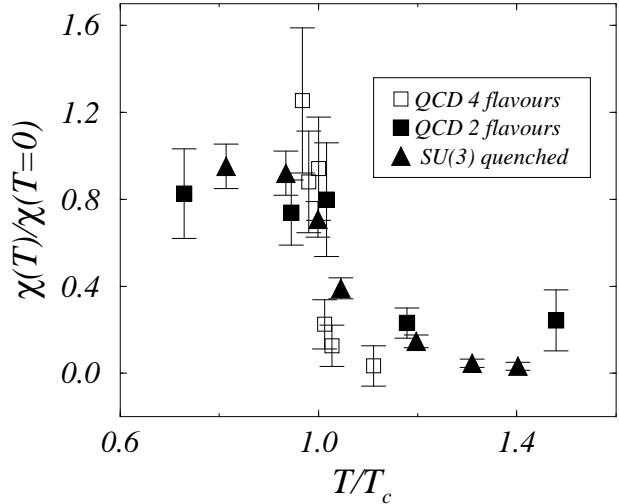}
\end{center}
\caption{\small{$\chi(T)/\chi(T=0)$ versus $T/T_c$ 
for $N_f=4$, $N_f=2$ and $N_f=0$.}}
\label{fig:wholeTNf4Nf2Nf0}
\end{figure}

The behaviour of the ratio
$\chi(T)/\chi(T=0)$ as a function of $T/T_c$ is shown in
Fig.~{\ref{fig:wholeTNf4Nf2Nf0}}. The case $N_f=2$ will
be discussed in the next section. For $N_f=4$ 
the value at zero temperature $\chi(T=0)$ is computed as the average
of all results at $T\leq T_c$. For comparison, we have included in the
figure the data for the quenched theory taken from Ref.~\cite{npb494}.
There is clear evidence that the topological
susceptibility in full QCD with $N_f=4$ presents a drop
at $T_c$. Apparently the drop is sharper than in the case of pure
Yang--Mills. However, the two theories have been simulated at rather
different lattice volumes ($32^3\times 8$ in the pure Yang-Mills case). 
Moreover, at the small values of 
$\beta$ used in our present simulation, we expect violations of
scaling in the ratio $T/T_c$. As a consequence we prefer not to draw
definite conclusions about the relative slopes.

\section{Full QCD with $N_f=2$}
\label{sec:2}

{} Full QCD with two flavours of quarks has been simulated by using
staggered fermions of bare mass $am=0.0125$ and the usual Wilson
action for the pure gauge sector. The updating has been performed with
the R--type HMC algorithm~\cite{gottlieb}, each trajectory
consisting in 120 steps of $\Delta\tau=0.004$ units of molecular
dynamics time. We have simulated at 
$\beta$=5.40, 5.50, 5.55, 5.60 and 5.70 both at zero and finite
temperature.

The lattice volume was $32^3\times 8$ for
finite $T$ and $16^4$ for $T = 0$. According to Ref.~\cite{krasnitz} at this 
temporal size 
the deconfinement transition occurs at $\beta_c=5.54(2)$. We have
checked this number by studying the Polyakov loop in a similar way
as in section 3.

We have checked that the sets of configurations produced during the 
simulation are well decorrelated and the distribution of topological
charge is well sampled. For the
case of $\beta=5.70$ we needed a longer simulation to
obtain a well sampled distribution of topological charge.

\begin{table}[hbtp]
\setlength{\tabcolsep}{1.88pc}
\centering
\caption{$\chi$, $T/T_c$ and $a$ versus $\beta$
for $N_f=2$ on a $32^3\times 8$ lattice.}
\begin{tabular}{cccll}
\hline
\hline
$\beta$ & $a$/fm & $T/T_c$ & $10^{-8}\times\chi$/MeV$^4$ \\
\hline
5.40 & 0.1742(50) & 0.73 & 5.16(1.65) \\
5.50 & 0.1343(72) & 0.94 & 6.21(1.74)  \\
5.55 & 0.1249(15) & 1.02 & 4.75(1.69)  \\
5.60 & 0.1076(27) & 1.18 & 2.27(0.72)  \\
5.70 & 0.0858(35) & 1.48 & 1.75(1.01)  \\
\hline
\hline
\label{tab:nf2}
\end{tabular}
\end{table}

In Table~{\ref{tab:nf2}} we report the data obtained from the
simulation at finite $T$. From the value extrapolated 
at $\beta_c= 5.54$, we infer the critical temperature
to be $T_c=194(10)(15)$ MeV where the first error is our statistical
and the second one the error induced by the indetermination in the
value of $\beta_c$.

In Fig.~{\ref{fig:wholeTNf4Nf2Nf0}} the behaviour of the ratio
$\chi(T)/\chi(T=0)$ as a function of $T/T_c$ is shown. For $N_f=2$
the value at zero temperature $\chi(T=0)$ has been obtained from an
independent simulation on symmetric lattices $16^4$ at the same values
of $\beta$ (see Table~{\ref{tab:nf2mass}}). 
There is a clear drop in the signal when the deconfinement
temperature is crossed. 

\begin{figure}[tbh]
\begin{center}
\includegraphics[width=0.5\textwidth]{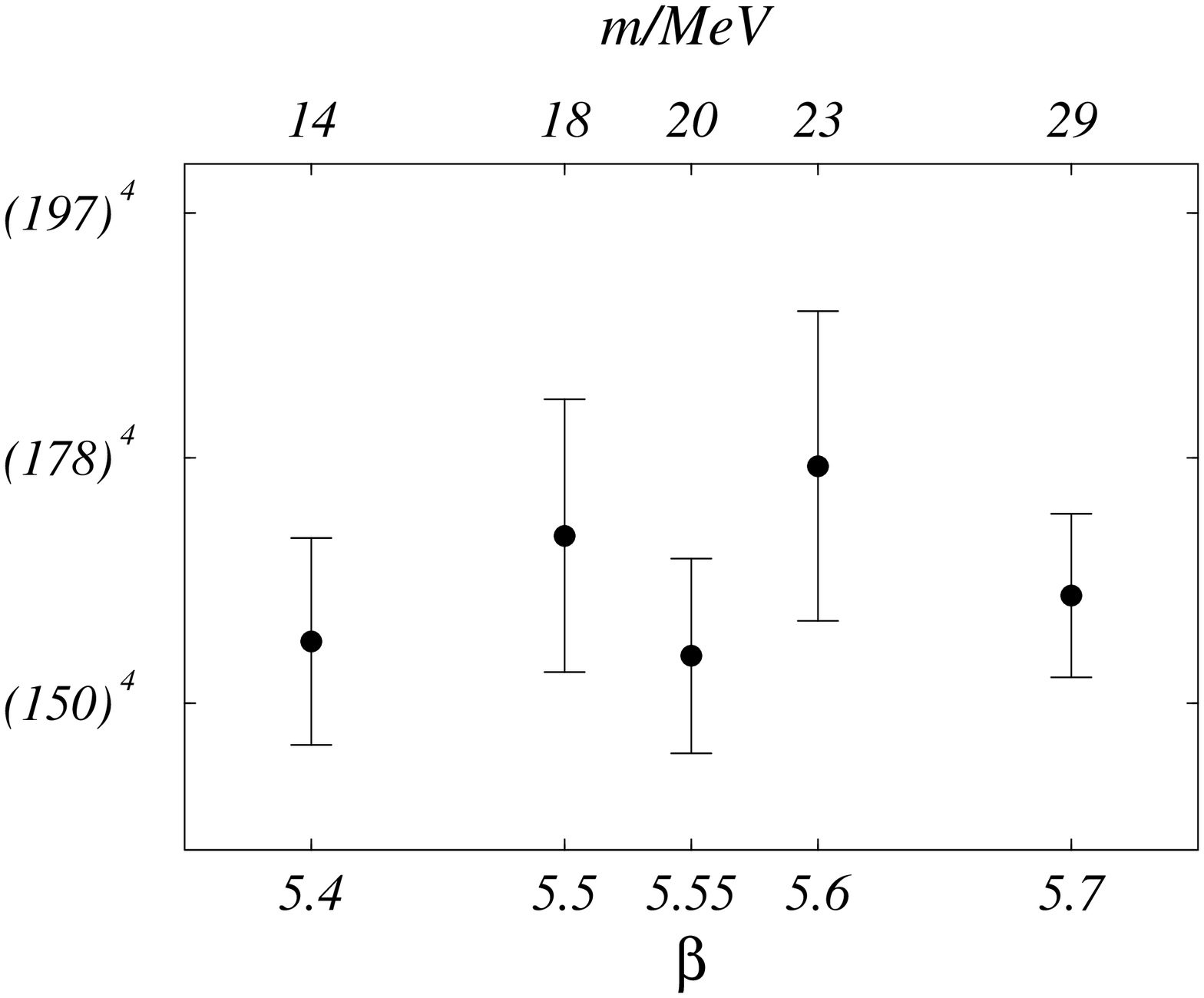}
\end{center}
\caption{\small{Dependence of the topological susceptibility on the
bare quark mass for $N_f=2$ at zero temperature.}}
\label{fig:massNf2}
\end{figure}

In Fig.~{\ref{fig:massNf2}} we plot the topological susceptibility
obtained at zero temperature for the five values of $\beta$. The corresponding
 set of data is listed in Table~{\ref{tab:nf2mass}}. Having fixed
$am$, the values of the bare mass vary as shown in 
the upper horizontal scale.
A fit with a constant value of the topological susceptibility gives
$\chi = (163 \pm 6)^4 \; {\rm MeV}^4$, with the statistical test
$\hbox{(chi--square)}/{\rm d.o.f} = 0.37$; a fit
with a linear homogeneous dependence on $m$, like Eq.~(\ref{eq:psibarpsi}), 
gives $\langle \overline{\psi} \psi \rangle_{\rm BARE} = (6.2 \pm 0.8) \times
10^7 \; {\rm MeV}^3$ and $\hbox{(chi--square)}/{\rm d.o.f} = 0.94$. 
Therefore our present 
data are compatible with both behaviours, and we are not able, 
within our errors, to check Eq.~(\ref{eq:psibarpsi}). 

\begin{table}[hbtp]
\setlength{\tabcolsep}{1.88pc}
\centering
\caption{$\chi$ versus $\beta$
for $N_f=2$ on a $16^4$ lattice.}
\begin{tabular}{cll}
\hline
\hline
$\beta$ & $10^{-8}\times\chi$/MeV$^4$ \\
\hline
5.40 & 6.25(2.11) \\
5.50 & 8.41(2.78) \\
5.55 & 5.96(1.99) \\
5.60 & 9.83(3.16) \\
5.70 & 7.19(1.67) \\
\hline
\hline
\label{tab:nf2mass}
\end{tabular}
\end{table}

\section{Conclusions}
\label{sec:conclusions}

We have studied the topological susceptibility $\chi$ in full QCD 
with 2 and 4 flavours of dynamical fermions.
At zero temperature, we have not enough precision to
check the chiral limit, Eq.~(\ref{eq:psibarpsi}).

At finite temperature, $\chi$
stays constant for values of $T$ up to the  deconfinement temperature $T_c$. 
At $T_c$ it presents a sudden drop and $\chi$ becomes compatible with
zero at $T \simeq 1.5\,T_c$ for $N_f=2$ and  $T \simeq 1.2\,T_c$ for $N_f=4$. 
This behaviour is qualitatively similar to that found
in Ref.~\cite{npb494} for the quenched case.

\section*{Acknowledgements}

B.A. thanks the Theory Group in Pisa for the kind hospitality.
{} Financial support from EC, Contract FMRX--CT97--0122, and from
MURST is acknowledged.

\newpage


\end{document}